\documentclass[reprint,aps,onecolumn,notitlepage]{revtex4-1}
\usepackage{graphicx}
\usepackage{amsmath,amssymb,mathbbol}
\usepackage{hyperref}
\hypersetup{
  colorlinks   = true, 
  urlcolor     = gray, 
  linkcolor    = red, 
  citecolor   = blue 
}
\usepackage{comment,subfigure,float}
\usepackage{verbatim}
\usepackage{courier}
\usepackage{array}
\newcolumntype{x}[1]{>{\centering\arraybackslash\hspace{0pt}}p{#1}}
\usepackage{graphicx}
\usepackage{xcolor}
\usepackage{bm,pgffor}
\usepackage{tcolorbox}
\usepackage{color}
\usepackage{dcolumn}
\usepackage{siunitx}

\newcommand{\highlight}[1]{\textcolor{red}{\textbf{#1}}}
\newcommand{\highlighta}[1]{\textcolor{blue}{\textbf{#1}}}
\begin{document}
\title{Simulating dense granular suspension rheology using LAMMPS}
\author{Christopher Ness}
\affiliation{School of Engineering, University of Edinburgh, Edinburgh, EH9 3JL, United Kingdom}

\begin{abstract}
Dense suspensions are widespread in nature, manufacturing and process engineering.
Particle-based simulations have proven to be an invaluable complement to experimental rheological characterisation,
serving as a virtual rheometer that enables rapid exploration of parameter space
and detailed scrutiny of microscopic dynamics.
To maximise the utility of such simulations,
it can be advantageous to exploit pre-existing, well-optimised, well-documented codes.
Here we provide a simple description of how to use LAMMPS to study the rheology of dense, granular suspensions.
\end{abstract}

\maketitle

\section{Introduction}
Suspensions of micron-sized particles with solid and liquid mixed in roughly equal proportion present intriguing flow properties that challenge physicists and engineers of all kinds~\cite{ness2022physics}.
A useful starting point for characterising their rheology is to understand the rate-independent behaviour (relevant for solid particles of $\approx10-1000\mu $m size),
demonstrated experimentally by~\citet{boyer2011unifying} and later reviewed by many others~(e.g.~\cite{guazzelli_pouliquen_2018}).
A number of particle-based simulations reproduce the rate-independent rheology~(e.g.~\cite{gallier_lemaire_peters_lobry_2014,cheal2018rheology}),
providing
(i) corroboration of the experimental result;
(ii) a source of particle-resolved data inaccessible experimentally;
and
(iii) a platform for examining systematically more complex microphysics,
for instance particle inertia~\cite{trulsson2012transition}, stress-induced friction~\cite{seto2013discontinuous} and adhesion~\cite{singh2019yielding}.
Many simulation techniques and codes are available, and we do not review these here.
In what follows, we describe how to generate numerical rheology data consistent with the rate-independent result using LAMMPS~\cite{plimpton1995fast,lammps}.

The simulation technique shares many details with traditional molecular dynamics,
and the reader is expected to be familiar with basic concepts including contact detection, neighbour listing, timestepping and so on.
Our specific model is more commonly labelled as a `discrete element method' owing to its similarity to approaches used in granular physics,
specifically the absence of thermal forces and the inclusion of particle-particle friction (both appropriate for the size range mentioned above).
The basic approach for obtaining a numerical rheology measurement is to
(i) initialise the system with a packing of non-overlapping spherical particles of desired size distribution at a desired volume fraction $\phi$;
(ii) evaluate the trajectory of each particle $i$ by numerically solving ``$F=ma$'' in the presence of a prescribed background fluid velocity gradient $\nabla\bm{u}^\infty$
and a set of pairwise hydrodynamic and contact interactions.
(We assume $\phi\gtrsim0.4$ throughout, otherwise a more detailed account of the hydrodynamics is required.)
When desired, a bulk stress tensor $\mathbb{\Sigma}$ is calculated from the interaction forces and particle positions,
thus generating rheology data \emph{viz.} the stress $\mathbb{\Sigma}$ as a function of deformation rate $\mathbb{E}$ (with $\mathbb{E}\equiv\frac{1}{2}(\nabla\bm{u}^\infty + {\nabla\bm{u}^\infty}^\mathrm{T}$)) and volume fraction~$\phi$.

\section{Dimensional analysis for rate-independent suspensions}
It is instructive to consider first a dimensional analysis.
This introduces the quantities (and their units) that we will define when setting up a simulation.
We consider a dense, granular (so we omit $k_bT$ from the following) suspension of stiff spheres under a flow with imposed deformation rate.
The principle particle properties (these set the length, mass and time scales) are
the characteristic particle radius $a$ [length],
the particle density $\rho$ [mass/length$^3$]
(taken throughout to be equal to the fluid density),
and the particle normal stiffness $k_n$ [mass/time$^2$] (this has a tangential counterpart $k_t$).
With respect to these quantities, 1 time unit corresponds to the inverse frequency of a mass $\rho a^3=1$ on a linear spring with stiffness $k_n=1$.
The remaining material properties to be defined are
the fluid viscosity $\eta_\mathrm{f}$ [mass/(length$\times$time)]
and
the particle-particle friction coefficient $\mu$ [dimensionless],
relevant for micron sized (and larger) particles.
The relevant macroscopic quantities are the size of the simulation box $L$ [length] and the volume fraction $\phi$ [dimensionless] therein.
The background fluid flow is characterised by a velocity field $\bm{u}^\infty$ [length/time] and its gradient (a tensor) $\nabla{\bm u}^\infty$ [1/time] (that we specify, and take to be spatially uniform), the time $t$ for which the flow was applied, and a stress tensor $\mathbb{\Sigma}$ [mass/(time$^2\times$length)] (that we measure).
We write a scalar velocity gradient as $\dot{\gamma}$ ($\equiv\partial u_x/\partial y$) and a scalar stress as $\Sigma_{xy}$ (the $xy$ component of $\mathbb{\Sigma}$).
A list (others are possible) of nondimensional parameters necessary to fully define a given suspension under given flow conditions is then: 
\begin{center}
\begin{tabular}{x{1.8cm} x{3.3cm} x{2.5cm} x{1.5cm} x{1.5cm} x{3.3cm} x{1.8cm}}
 (i) $a/L$&
 (ii)  $\dot{\gamma} \sqrt{\rho a^3/k_n}$&
 (iii) $\rho\dot{\gamma}a^2/\eta_\mathrm{f}$&
(iv) $\mu$&
(v) $\phi$&
(vi) $\eta_r \equiv \Sigma_{xy}/\eta_\mathrm{f}\dot{\gamma}$&
(vii) $\dot{\gamma}t$
\end{tabular}
\end{center}
Setting (i)-(iii) to be $\ll 1$ ensures, respectively, bulk conditions, stiff particles and no particle inertia.
Under these conditions,
and assuming $\mu$ is constant (i.e. particle friction is Coulombic)
and we shear to steady state ($\dot{\gamma}t\to\infty$),
we have simply that $\eta_r=\eta_r(\phi)$, hence the label `rate-independent'.
This is consistent with the result of~\citet{boyer2011unifying}, and will be the focus of our example below.

\section{Particle-level forces and shearing}

\begin{figure}[b]
\vspace{3mm}
\includegraphics[trim = 0mm 38mm 135mm 0mm, clip,width=0.9\textwidth,page=1]{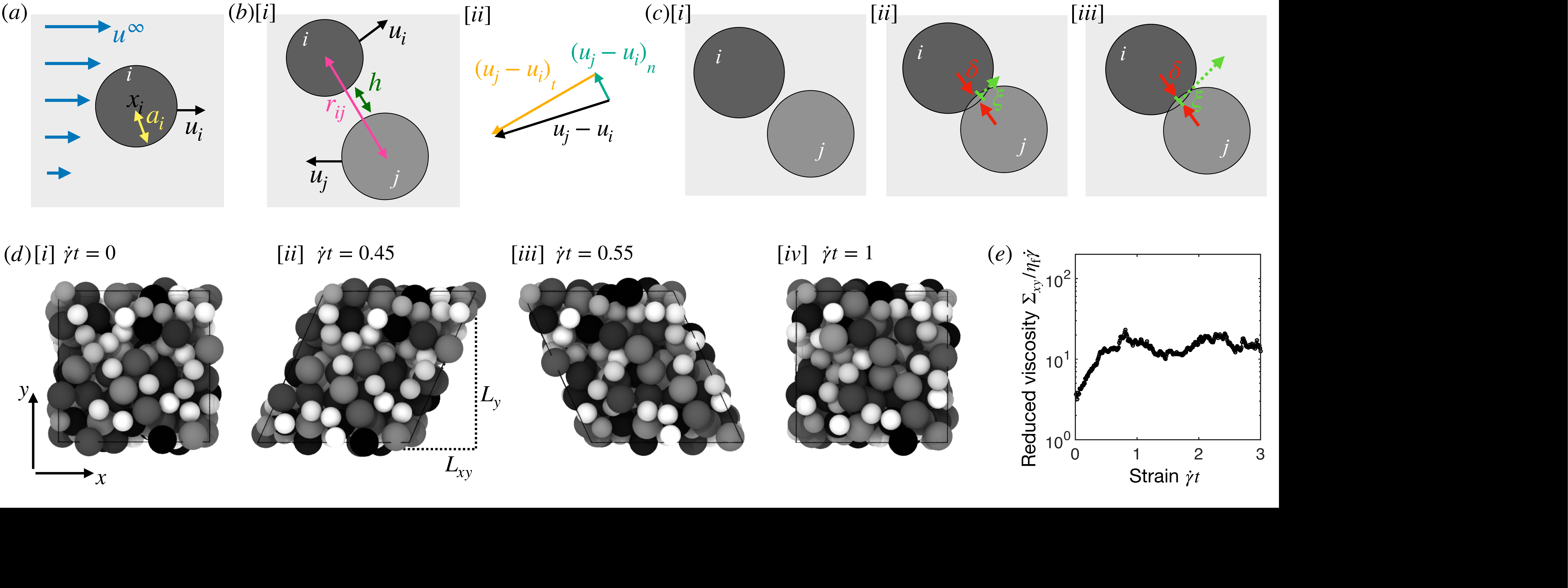}
\caption{
Particle-level physics in a simplified dense suspension.
(a) A particle (radius $a_i$, position $\bm{x}_i$, velocity $\bm{u}_i$) in a fluid with streaming velocity $\bm{u}^\infty$;
(b) Pairwise lubrication interaction showing
[i] particle velocities $\bm{u}_i$, $\bm{u}_j$, centre-to-centre vector $\bm{r}_{ij}$ and surface separation $h$;
[ii] relative velocity $(\bm{u}_j-\bm{u}_i)$ and its components normal $(\bm{u}_j-\bm{u}_i)_n$ and tangential $(\bm{u}_j-\bm{u}_i)_t$ to $\bm{r}_{ij}$;
(c) Contact force showing
[i] particles approaching;
[ii] particles entering contact with overlap $\delta$ and initiation of sliding with $\bm{\xi}=0$;
[iii] contact with overlap $\delta$ and accumulated, nonzero, tangential stretch $\bm{\xi}$.
(d) Subjecting the suspension to a steady simple shear. Shown are increments of the strain ($\dot{\gamma}t=0$ [i], 0.45 [ii], 0.55 [iii], 1 [iv]) illustrating the remapping procedure used by LAMMPS.
(e) Reduced viscosity $\Sigma_{xy}/\eta_\mathrm{f}\dot{\gamma}$ as a function of the accululated strain $\dot{\gamma}t$.
}
\label{figure1}
\end{figure}

Rate-independent rheology is obtained by subjecting particles to three types of force and torque: Stokes drag, pairwise lubrication and pairwise contact.
The full form of these is reported by several authors~\cite{trulsson2021transition,mari2014shear,cheal2018rheology,ge2020implementation} and need not be repeated here.
Instead we describe the forces in simplified terms.

The Stokes drag (Fig.~\ref{figure1}(a)) on particle $i$
(radius $a_i$)
is proportional to the difference between its velocity $\bm{u}_i$ and the fluid streaming velocity at its centre $\bm{u}^\infty(\bm{x}_i)$:
\begin{equation}
\bm{F}^{d}_i = -6\pi\eta_\mathrm{f}a_i(\bm{u}_i - \bm{u}^\infty(\bm{x}_i)) \text{.}
\label{eq:drag}
\end{equation}
This force is essentially what induces flow in the simulation, causing particles to conform to the streaming velocity set by $\bm{u}^\infty$.
Similarly, a torque acts to cause the particles to rotate with angular velocity set by $\frac{1}{2}(\nabla\times\bm{u}^\infty)$.
Neighbouring particles $i$ and $j$ with centre-to-centre vector $\bm{r}_{i,j}$ (Fig.~\ref{figure1}(b)[i]) experience lubrication forces (see~\cite{kim1991,jeffrey1992calculation}) dependent on the gap $h$ between them
and their
relative velocity (Fig.~\ref{figure1}(b)[ii]).
The leading term of the force on particle $i$ (assuming it has equal radius to particle $j$) scales with $1/h$ and the normal component of the pairwise velocity difference:
\begin{equation}
\bm{F}^l_{i,j} = \frac{3}{2}\pi a_i^2 \eta_\mathrm{f}\frac{1}{h}(\bm{u}_j - \bm{u}_i)_n \text{.}
\end{equation}
These lubrication forces oppose relative motion between particle pairs.
They are prevented from diverging at contact by setting a lower limit on the allowed value of $h$ (typically $\mathcal{O}(10^{-3}a_i)$).
A torque also acts to resist relative rotation between $i$ and $j$.
Overlapping particle pair $i$ and $j$ (Fig.~\ref{figure1}(c)[i]) experience repulsive contact forces dependent upon the scalar overlap $\delta$ (Fig.~\ref{figure1}(c)[ii]) and the tangential displacement accumulated over the duration of the contact $\bm{\xi}$ (Fig.~\ref{figure1}(c)[iii]):
\begin{equation}
\bm{F}^c_{i,j} = k_n\delta \bm{r}_{i,j}/|\bm{r}_{i,j}| - k_t\bm{\xi} \text{.}
\label{eq:contact}
\end{equation}
The friction coefficient $\mu$ sets an upper bound on $\bm{\xi}$ through $|\bm{\xi}|\leq\mu k_n\delta/k_t$.

The stress contribution from drag forces is proportional to $\mathbb{E}$.
The $\alpha,\beta$ component of the stress due to lubrication and contact is found, respectively, by summing $(F^{l,\alpha}_{i,j}r^\beta_{i,j} + F^{l,\beta}_{i,j}r^\alpha_{i,j})/2$ and $F^{c,\alpha}_{i,j}r^\beta_{i,j}$ over all pairs.
The forces are summed on each particle and the trajectories are then updated 
according to Newtonian dynamics, using a numerical scheme with timestep chosen to be small compared to $\sqrt{\rho a^3/k_n}$ and $\rho a^2/\eta_\mathrm{f}$.

In LAMMPS the simulation box deforms according to the specified $\nabla{\bm u}^\infty$.
For instance, when the only nonzero element of $\nabla{\bm u}^\infty$ is an off-diagonal, shearing is applied by tilting the \emph{triclinic} box (at fixed volume) according to
$L_\text{xy}(t) = L_\text{xy}(t_0) + L_y\dot{\gamma}t$, Fig.~\ref{figure1}(d).
When the strain ($\gamma=\dot{\gamma}t$) reaches 0.5 in this example, the system is remapped to a strain of -0.5. This has no effect on the particle-particle forces or on the stress, and is simply a numerical tool to permit unbounded shear deformation while preventing the domain from becoming elongated in one axis.

\section{LAMMPS inputs and outputs}
The above physics are implemented in 
LAMMPS (Large-scale Atomic/Molecular Massively Parallel Simulator)~\cite{plimpton1995fast},
a classical molecular dynamics code written in C\texttt{++}.
The LAMMPS documentation should be referred to at all times~\cite{lammps}.
A skeletal set of instructions for downloading and compiling LAMMPS
and running the scripts below is provided at~Ref~\cite{ness}.
Our strategy is to separate the generation and shearing of suspensions into two distinct simulations.
The first defines dimensionless numbers (i) and (v) above, producing non-overlapping particle packings in a cuboidal, periodic domain of fixed $\phi$;
the second defines (ii), (iii), (iv) and (vii) then applies the deformation allowing us to measure (vi).
This decomposition allows one to build a library of configurations at different $\phi$ that can be reused many times for different deformation protocols.
LAMMPS takes as its input a text file containing a list of commands and their arguments.
In the following we don't describe each of these in detail,
but instead provide minimal scripts, indicating where the physics above enters.
Each of the commands is fully described in the LAMMPS documentation~\cite{lammps}.
%

\begin{center}
\begin{tcolorbox}[title=\texttt{in.create},leftrule=2mm,toprule=0mm,width=0.8\textwidth]
\footnotesize{
\texttt{\# SETTINGS\\
atom\_style sphere\\
comm\_modify mode single vel yes\\
\\
\# GENERATE A CUBIC, PERIODIC SIMULATION BOX\\
boundary p p p\\
region reg prism 0 \highlight{14.6381} 0 \highlight{14.6381} 0 \highlight{14.6381} 0 0 0 units box\\
create\_box 2 reg\\
\\
\# GENERATE THE PARTICLES AND SPECIFY THEIR SIZE\\
create\_atoms 1 random \highlight{100} \highlighta{123456} NULL\\
create\_atoms 2 random \highlight{100} \highlighta{123457} NULL\\
set type 1 diameter \highlight{2.8}\\
set type 2 diameter \highlight{2}\\
set type 1 density \highlight{1}\\
set type 2 density \highlight{1}\\
\\
\# SPECIFY THE PARTICLE-PARTICLE INTERACTION\\
pair\_style granular\\
pair\_coeff * * hooke \highlight{10000} 0 tangential linear\_history \highlight{7000} 0 \highlight{0.1}\\
\\
\# SPECIFY THE OUTPUTS\\
thermo 10000\\
dump	id all custom 10000 create.dump id x y z radius\\
log create.log\\
\\
\# SPECIFY THE TIMESTEP, THE INTEGRATION SCHEME AND RUN\\
timestep \highlight{0.0001}\\
fix 1 all nve/sphere\\
fix     2 all viscous \highlight{1.88} scale 1 1.4\\
run \highlight{1000000}\\
\\
\# WRITE AN OUTPUT FILE\\
write\_data data.file
}}
\end{tcolorbox}\end{center}

\paragraph*{Creating particle packings.--}
For the time being, it isn't necessary to specify the `full' physics described above, 
nor do we need to output the stresses.
Rather, we need just enough detail to create assemblies of non-overlapping particles.
We therefore omit lubrication forces at this stage for simplicity.
We first generate particles (of two types, each having a different radius) with random coordinates in a box of set dimensions.
Their overlaps generate contact forces (following Eq.~\ref{eq:contact}) that lead to motion;
damping against a stationary background fluid (i.e. $\bm{u}^\infty=0$ in Eq.~\ref{eq:drag}) extracts energy until the system comes to rest.
The properties of the configuration of particles produced by this script don't really matter:
we are not trying to sample an `equilibrium' configuration (this is not relevant for granular systems) but simply create a packing with no (or minimal) overlaps.
Shown in the \texttt{in.create} panel is an example input script (`\#' indicates comments) to generate a suspension with $\phi=0.5$.

Highlighted in red from top to bottom are (with units as stated earlier):
(i) the size of the cubic simulation box $L = 14.6381$;
(ii) the numbers and radii of particles of types 1 and 2: $N_1=N_2=100$,
$2a_1 = 2$ and $2a_2=2.8$,
thus setting the volume fraction as $\phi={4/3}\pi(N_1 a_1^3 + N_2 a_2^3)/L^3=0.5$;
(iii) the particle density $\rho$;
(iv) the particle stiffness $k_n=10000$ (with $k_t=7k_n/10$);
(v) the particle-particle friction coeffient $\mu=0.1$;
(vi) the timestep;
(vii) the fluid viscosity $\eta_\mathrm{f}=0.1$ (the number we put in the script is $6\pi\eta_\mathrm{f}a$);
(viii) the number of timesteps to run.
Highlighted in blue are the seeds used to generate the initial particle positions.
New realisations can be generated by rerunning the simulation with different numbers here.
This script produces \texttt{data.file}, containing a snapshot of the system after the final timestep, to be read in by future scripts.
The file contains a list of the particle ID, diameter, density, coordinates ($x$, $y$, $z$) and velocity components. 
It also produces a dump file (create.dump) that can be visualised using e.g. Ovito~\cite{stukowski2009visualization}.

\paragraph*{Shearing particle packings.--}
The second script (\texttt{in.run}, see panel below) takes \texttt{data.file} as an input and applies a deformation to the sample.
The key inputs to this script are the parameters related to the particle-particle interaction ($k_n$, $k_t$, $\mu$ and $\eta_\mathrm{f}$)
and the shear rate, set by specifying the components of $\nabla\bm{u}^\infty$.
Through these we set the values of dimensionless control parameters (ii), (iii), (iv) and (vii) listed above.
The remaining content of the input script is concerned with specifying the bulk stress calculation and requesting it as an output, necessary for obtaining (vi).
\begin{center}
\begin{tcolorbox}[title=in.run,leftrule=2mm,toprule=0mm,width=0.8\textwidth]
\footnotesize{
\texttt{\# SETTINGS\\
atom\_style	sphere\\
comm\_modify	mode single vel yes\\
newton off\\
\\
\# READ THE PARTICLE CONFIGURATION\\
read\_data data.file\\
\\
\# SPECIFY THE PARTICLE-PARTICLE INTERACTION\\
pair\_style hybrid/overlay granular lubricate/bmpoly \highlight{0.1} 1 1 0.001 0.05 1 0\\
pair\_coeff * * granular hooke \highlight{10000} 0 tangential linear\_history \highlight{7000} 0 \highlight{0.1}\\
pair\_coeff * * lubricate/bmpoly\\
\\
\# DO THE STRESS CALC\\
compute		str all pressure NULL pair\\
\\
\# SPECIFY THE OUTPUTS\\
thermo\_style custom time c\_str[1] c\_str[2] c\_str[3] c\_str[4] c\_str[5] c\_str[6]\\
thermo 10000\\
dump	id all custom 10000 run.dump id x y z radius\\
log run.log\\
\\
\# SPECIFY THE TIMESTEP, THE INTEGRATION SCHEME AND RUN\\
timestep	0.0001\\
fix		1 all nve/sphere\\
fix 2 all deform 1 xy erate \highlight{0.001} remap v\\
run 30000000
}}\end{tcolorbox}\end{center}
Highlighted from top to bottom (units as before) are
the fluid viscosity $\eta_\mathrm{f} = 0.1$,
the particle stiffness $k_n=10000$ ($k_t=7000$) and friction coefficient $\mu=0.1$
and the shear rate $\dot{\gamma}=0.001$.
Using the command \texttt{fix deform} we have specified just the $xy$ component of $\nabla\bm{u}^\infty$, $\dot{\gamma}$ as defined above. The other components are $0$ by default, so this leads to a simple shear with flow in $x$ and gradient in $y$ i.e. $\bm{u}^\infty = (\dot{\gamma}y,0,0)$.
The script runs at this $\dot{\gamma}$ for 30000000 timesteps, each of duration 0.0001. Thus the total shear strain is $\dot{\gamma}t = 0.001\times30000000\times0.0001 = 3$.
Note that the \texttt{fix viscous} command is not required here because the drag force is applied within the lubrication pair style.

\paragraph*{Outputs from shearing simulation.--}
There are two different types of output produced by LAMMPS during a simulation run, \emph{log} files and \emph{dump} files. 
Log files are typically described as containing thermodynamic data,
but for our purposes we can interpret this as bulk suspension or derived properties,
usually the components of the stress tensor~$\mathbb{\Sigma}$, but also e.g. the average particle contact number.
As specified in \texttt{in.run} above, the log file (\texttt{run.log}) contains the accumulated simulation time $t$, followed by the 6 unique components of the stress tensor in order ($xx$, $yy$, $zz$, $xy$, $xz$, $yz$), output every 10000 timesteps (specified by the \texttt{thermo} command):
\begin{center}
\begin{tcolorbox}[title=run.log,leftrule=2mm,toprule=0mm,colframe=red!50!black,width=0.8\textwidth]
\footnotesize{
\texttt{Time c\_str[1] c\_str[2] c\_str[3] c\_str[4] c\_str[5] c\_str[6] \\
 0              0.00073391 -0.0018116  -0.00028190 -0.001589    0.00018953 -0.00058983\\
 1              0.00063973  0.0014631   0.00086064 -8.3279e-05 -5.7537e-05 -0.00020223\\
 2              0.00040622  0.00045020  0.00066278 -0.00015159  9.5433e-05 -0.00012521\\
 3              0.00029818  0.00032181  9.1296e-05 -0.00044916  6.2575e-05 -0.00012202\\
$\dots$
}}\end{tcolorbox}\end{center}
These stress components have units [mass/(time$^2\times$length)]. In order to express in dimensionless units (the ``reduced viscosity'' as it is conventionally represented) for comparison to experimental data, one must divide by $\eta_\mathrm{f}\dot{\gamma}$.
Shown in Fig.~\ref{figure1}(e) is a plot of the reduced viscosity as a function of the accumulated strain.
The reduced viscosity is in this instance taken simply as the $xy$ component of the stress ($\Sigma_{xy}$, the 4th column of the stress outputs in the log file) divided by $\eta_\mathrm{f}\dot{\gamma}$ whereas the strain is the accumulated time $t$ multiplied by $\dot{\gamma}$. From the content of the log file one might also compute e.g. the viscous number $\eta_\mathrm{f}\dot{\gamma}/P$ (with $P$ the mean of the diagonal components of $\mathbb{\Sigma}$ ($\Sigma_{xx}$, $\Sigma_{yy}$, $\Sigma_{zz}$), the normal stress differences $\Sigma_{xx}-\Sigma_{yy}$, $\Sigma_{yy}-\Sigma_{zz}$ and so on.

Dump files contain particle-level information (positions, velocities, radii) or contact level information (forces, relative positions), usually output at fixed intervals.
In the script above this is specified (by the $\texttt{dump}$ command) to be every 10000 timesteps.
This example script produces \texttt{run.dump}, which lists the particle IDs, positions and radii at specified time intervals.
This file might be used for post-processing, for instance to compute structural properties, to follow particle trajectories or to be read directly into various visualisation packages (Fig.~\ref{figure1}(d) was generated using Ovito~\cite{stukowski2009visualization}, for instance).
\begin{center}
\begin{tcolorbox}[title=run.dump,leftrule=2mm,toprule=0mm,colframe=red!50!black,width=0.8\textwidth]
\footnotesize{
\texttt{ITEM: TIMESTEP\\
0\\
ITEM: NUMBER OF ATOMS\\
200\\
ITEM: BOX BOUNDS xy xz yz pp pp pp\\
0.0000000000000000e+00 1.4638100000000000e+01 0.0000000000000000e+00\\
0.0000000000000000e+00 1.4638100000000000e+01 0.0000000000000000e+00\\
0.0000000000000000e+00 1.4638100000000000e+01 0.0000000000000000e+00\\
ITEM: ATOMS id x y z radius \\
1 9.41798 1.04987 0.152439 1.4\\
2 9.03954 7.66538 5.31888 1.4\\
3 0.69716 1.35518 9.27356 1.4\\
$\dots$
}}\end{tcolorbox}\end{center}

By repeating this pair of simulations at a range of $\phi$ (achieved by changing the value of $L$ in \texttt{in.create}) and taking time (and ensemble) averages of the components of $\mathbb{\Sigma}$,
one can reproduce the rate-independent rheology of~\citet{boyer2011unifying}.
Moreover, one may relax the conditions specified in the dimensional analysis to explore additional physics:
(i) introducing particle inertia by increasing the value of $\rho\dot{\gamma}a^2/\eta_\mathrm{f}$;
(ii) introducing particle softness by increasing the value of $\dot{\gamma} \sqrt{\rho a^3/k_n}$,
and so on.

\section{Closing remarks}

We have provided a brief description of how to use the molecular dynamics code LAMMPS to generate dense, granular suspension rheology data.
Examples of the use of this approach to study the physics of dense suspensions can be found in Refs~\cite{ness2016two,ness2018shaken,guy2020testing,gillissen2020constitutive,gillissen2020modeling,niu2020tunable}.
Moreover, the flexibility of the code allows one to simulate more complex geometries, both within LAMMPS
~\cite{ramaswamy2017confinement,ness2017linking,blair2022shear}
and within derivative codes such as LIGGGHTS~\cite{cabiscol2021application}.

C.N. acknowledges support from the Royal Academy of Engineering under the Research Fellowship scheme,
and useful discussions many collaborators and students. Contact: chris.ness@ed.ac.uk.

\bibliography{library}

\end{document}